\documentclass{emulateapj}
\usepackage{times}

\def\la{\mathrel{\mathpalette\fun <}}
\def\ga{\mathrel{\mathpalette\fun >}}

\def\fun#1#2{\lower0.837ex\vbox{\baselineskip0ex\lineskip0.209ex
  \ialign{$\mathsurround=0ex#1\hfil##\hfil$\crcr#2\crcr\sim\crcr}}}

\newcommand{\et}{et al.\ }

\def\sles{\lower2pt\hbox{$\buildrel {\scriptstyle <}
   \over {\scriptstyle\sim}$}}
 
\def\sgreat{\lower2pt\hbox{$\buildrel {\scriptstyle >}
   \over {\scriptstyle\sim}$}}

\def\la{\mathrel{\mathpalette\fun <}}
\def\ga{\mathrel{\mathpalette\fun >}}

\begin{document}

\title{ Discovery of an Afterglow Extension of the Prompt 
Phase of Two Gamma Ray Bursts Observed by  {\em Swift} }

\shortauthors{Barthelmy \et}
\author{
S.~D.~Barthelmy\altaffilmark{1},
J.~K.~Cannizzo\altaffilmark{1,2},
N.~Gehrels\altaffilmark{1},
G.~Cusumano\altaffilmark{3},
V.~Mangano\altaffilmark{3},
P.~T.~O'Brien\altaffilmark{4},
S.~Vaughan\altaffilmark{4},
B.~Zhang \altaffilmark{5},
D.~N.~Burrows\altaffilmark{6},
S.~Campana\altaffilmark{7},
G.~Chincarini\altaffilmark{7,8},
M.~R.~Goad\altaffilmark{4},
C.~Kouveliotou\altaffilmark{9},
P.~Kumar\altaffilmark{10},
P.~M\'{e}sz\'{a}ros\altaffilmark{6},
J.~A.~Nousek\altaffilmark{6},
J.~P.~Osborne\altaffilmark{4},
A.~Panaitescu\altaffilmark{11},
J.~N.~Reeves\altaffilmark{1,12,13},
T.~Sakamoto\altaffilmark{1,14},
G.~Tagliaferri\altaffilmark{7},
R.~A.~M.~J.~Wijers\altaffilmark{15}
}
\altaffiltext{1}{NASA/Goddard Space Flight Center, Exploration of the Universe Division, Greenbelt, MD 20771}
\altaffiltext{2}{Joint Center for Astrophysics, University of Maryland, Baltimore County, 
                 Baltimore, MD 21250}
\altaffiltext{3}{INAF - Istituto di Astrofisica Spaziale e Cosmica, Via Ugo La Malfa 153, I-90146 Palermo,
                 Italy}
\altaffiltext{4}{Department of Physics \& Astronomy, University of Leicester, Leicester, LE1 7RH,  UK}
\altaffiltext{5}{Department of Physics, University of Nevada Las Vegas, Las Vegas, NV 89154-4002}
\altaffiltext{6}{Department of Astronomy \& Astrophysics, Department of Physics,
                   Pennsylvania State University, University Park, PA 16802}
\altaffiltext{7}{INAF - Osservatorio Astronomico di Brera, Via Bianchi 46, I-23807 Merate, Italy}
\altaffiltext{8}{Universit\'{a} degli studi di Milano-Bicocca, Dipartimento di Fisica, 
                 Piazza delle Scienze 3, I-20126  Milan, Italy}
\altaffiltext{9}{NASA/Marshall Space Flight Center, NSSTC, XD-12, 320 Sparkman Dr., Huntsville, AL 35805}
\altaffiltext{10}{Department of Astronomy, University of Texas at Austin, Austin, TX, 78712}
\altaffiltext{11}{Space Science and Applications, MS D466, Los Alamos National Laboratory, Los Alamos, NM 87545}
\altaffiltext{12}{Department of Physics \& Astronomy, Johns Hopkins University, Baltimore, Md 21218} 
\altaffiltext{13}{Universities Space Research Association, 10211 Wincopin Cir., Suite 500, 
                  Columbia, Md, 21044-3432}
\altaffiltext{14}{National Research Council, 2101 Constitution Ave NW, Washington, DC 20418}
\altaffiltext{15}{Astronomical Institute ``Anton Pannekoek'', University of Amsterdam, Amsterdam, 
                   the Netherlands}

\begin{abstract}

                 BAT and XRT 
 observations of two recent well-covered GRBs observed by {\em Swift},
  GRB 050315 and GRB 050319, show 
clearly 
    a prompt component
   joining the onset 
    of the afterglow 
emission. 
    By fitting a 
 power law 
 form to the $\gamma-$ray spectrum,
  we extrapolate  the time dependent fluxes measured by the
BAT, in the energy band $15-350$ keV,
     into the spectral regime observed by the XRT
  $0.2-10$ keV, and examine the functional form of the rate of
decay of the two light curves.
    We find that the BAT and XRT light curves merge to form a  unified
    curve. There is a period of steep decay up to $\sim300$ s, followed
by a flatter decay.
    The duration of the steep decay, $\sim100$ s 
            in the source
frame after correcting for cosmological time dilation,
    agrees roughly with a theoretical estimate for the deceleration time
 of the relativistic ejecta as it interacts with circumstellar material.
   For GRB 050315,
the steep decay can  be characterized by an exponential form,
where one $e-$folding decay time
  $\tau_e$(BAT)$\simeq24\pm2$ s, and $\tau_e$(XRT)$\simeq35\pm2$ s.
   For GRB 050319,
  a power law 
   decay    $- d\ln f/d\ln t = n$, where $n\simeq3$, provides a reasonable fit.
  The early time X-ray fluxes are consistent with 
  representing the lower energy tail of the prompt emission, 
   and provide our first quantitative  measure of 
the decay of the prompt $\gamma-$ray emission over a large dynamic range
  in flux.
   The initial steep decay 
    is expected due to the delayed high latitude photons
   from a curved shell of
   relativistic plasma
          illuminated only for a  short interval.
%
%
%
%
The overall conclusion is that the prompt phase of
GRBs remains observable for hundreds of seconds longer than
previously thought.

\end{abstract}


 \section{                  Background }

Gamma-Ray Bursts (GRBs) are 
 among the most energetic phenomena in the Universe,
    and are believed to contain gas with the
  highest  bulk-flow  Lorentz factors.
 GRBs belonging to the ``long'' class, with duration $\ga 2$ s
           (Kouveliotou et al. 1993),
      are thought to herald the death of a massive star
possessing high angular momentum, with the additional constraint
 that our line of sight coincides almost exactly with the
rotational axis of the progenitor star.
The apparent isotropic 
equivalent energies of $\sim3\times 10^{53}$ erg
   decrease  to $\sim5\times 10^{50}$ erg when one corrects
for   beaming (Frail et al. 2001, see also Panaitescu \& Kumar 2001). 
%
 The prompt emission from GRBs is thought
to come from a relativistically expanding fireball 
(Rees \& M\'esz\'aros 1992, 1994),
likely ejected during the collapse of massive stars
(MacFadyen \& Woosley 1999).
Because of the 
  traditionally 
      long delay between the observations of
          the GRB prompt emission
 and the start of the  afterglow observations, 
       the exact site of
the prompt emission has remained largely unknown. It has been argued that
it                  could either come from the internal shocks (Rees
\& M\'esz\'aros 1994) or from the external shocks
             (see, e.g., Zhang \& M\'esz\'aros
2004 and references therein).
      If the prompt emission were due to
external shocks, one  would see a continuous variation in flux between the prompt
and afterglow light curves, with the decay  slopes being equal.
   If it  were caused by
internal shocks, one should expect distinct components for the
 $\gamma-$ray light curves
 and the late afterglow. Looking for the bridge between the 
      early, $\gamma-$ray light curve ($\la 100$ s)
 and the later, X-ray  light curve ($\ga 100$ s)  
            is therefore essential in clarifying
the emission site for the early flux.
           The unique capability of {\em Swift} makes
this possible. 
 In particular,    early-time XRT data reveal
 that early X-ray afterglow shows a distinct steeply decaying component
 followed by a shallower, more standard decaying component
   (Tagliaferri et al. 2005, 
  Nousek et al. 2005).

    The finite $\gamma-$ray background of large
FOV detectors such as BATSE limits the available dynamic range in flux
to about two  orders of magnitude, except for unusually bright GRBs.
  For instance, Giblin et al. (1999) examined the BATSE decay light curve
of GRB 980923 and fit a decay law of the form $A(t-t_0)^{-n}$, where
$n = 1.8\pm0.02$. Other workers have carried out 
  similar studies and placed constraints on the decay index:
       $n$(GRB 920723)$ = 0.69 \pm 0.17$ 
(Burenin et al. 1999),   
     $n$(GRB 910402)$= 0.7$ and
     $n$(GRB 920723)$= 0.6$
(Tkachenko et al. 2000),
     and 
   $n$(GRB 990510)$= 3.7$
(Pian et al. 2001).
  Also, in't Zand et al. (2001)
 found a steep fall-off of the $2-10$ keV
 emission of GRB 010222 after 100 s.

%
%
%
   Connaughton (2002) co-added the background-subtracted BATSE light curves
for 400 long GRBs, and found $n\simeq 0.6$ for the ensemble decay.
 It is not clear how physically  meaningful this averaged value is, given the potential
variety of decays for different bursts, and the systematics of the background
subtraction for individual bursts.
 A related issue is that of how to ``line up'' different GRBs,
  i.e., the choice of $t_0$.
   For instance, if each distinct spike within a multi-spike GRB 
  results from
a       $\delta$-function injection of energy into a relativistic plasma, 
   the relevant $t_0$  for times well past the end of the GRB would be the
starting time for the last spike. The use of a physically inappropriate $t_0$
    would smear out the results of an ensemble average.
  There may also be a dependence of the results on the energy range being  utilized.

{\em Swift} was launched into a low-Earth orbit on 20 Nov 2004
(Gehrels et al. 2004).  It contains three instruments, the Burst Alert
Telescope (BAT; Barthelmy et al.  2005) with an energy range of $15
-350$ keV, the X-Ray Telescope (XRT; Burrows et al. 2005a) with an
energy range of $0.3 -10$ keV, and the UV/Optical Telescope (UVOT;
Roming et al. 2005) with a wavelength range of $170 - 650$ nm.  The
BAT initially detects the GRB and transmits a $1-3$ arc-min position
to the ground within $\sim12-45$ s.  The spacecraft then autonomously
slews to the GRB location within 20 to 75 s, at which time
observations with the two narrow-field instruments XRT and UVOT begin.

 For this study we consider two of 
     the best cases with known redshifts
   $-$
  GRB 050315 and GRB 050319.
    These are also the longer of the
   long bursts and so potentially allow us to test
   the relation between BAT and XRT fluxes during the
   near-overlap time of useful data with the two instruments.
%

\section{Data Analysis}

  The BAT data analysis is performed using the 
  {\em Swift} software  package (HEAsoft 6.0). 
  From the known GRB position
 determined from the initial trigger,
 the shadow mask weighting pattern for this position is
calculated for the coded aperture.
 The background is 
subtracted using the modulations of the coded aperture.
  In this technique, photons with energy greater than  150 keV become
transparent to the coded mask and are treated as a background.
The effective BAT energy range is from 15 keV to 150 keV in this
mask-weighted technique.

The BAT spectrum and the detector response matrix (DRM) are created
 using the                                         HEAsoft 6.0 software  packages,
and the {\em Swift} calibration database (CALDB 20050327).
 We also apply  an energy-dependent systematic error vector
 to the spectral files before doing spectral fitting.\footnote
{\tt http://legacy.gsfc.nasa.gov/docs/swift/analysis/\\
bat\_digest.html}
The background subtracted (mask-weighted) spectral data are used
in the analysis.
  The XSPEC v11.3.1 software package is used for fitting the data
to the model spectrum.

 {\em Swift} was slewing during GRB 050319, and the BAT trigger
is disabled during this  interval.
    The actual GRB began $\sim135$ s before the  originally reported
   trigger time $t_0$, which is now known to 
    represent the 
onset of the last of the 4 spikes comprising the GRB. 
   Nevertheless, in this study we utilize
  the original $t_0$ value, and restrict our attention
  only to the last spike.
Each individual spike would have a decay in X-rays
associated with it, and in any  given train of spikes 
           constituting the entire GRB, only the most
recent would be of relevance since the earlier ones would 
  largely have decayed by the later time.
  This convention for GRB 050319 concerning
   $t_0$ is 
        different than that adopted  
   by  Cusumano et al. (2005 = C05), 
   who took the trigger time for the first spike
    in the GRB 050319 complex.



\subsection{Methodology }

 We  calculate the decay of the  prompt emission as follows:
   We first extract the BAT light  curve in the energy range $15-350$ keV,
then fit a power law to the spectrum over the central 50\% of the 
fluence, i.e., $T_{50}$, then we extrapolate this emission  into the $0.2-10$ keV
energy range. The conversion factor for each GRB is calculated using 
 the flux calculator tool
{\tt PIMMS}.
 The power law index inferred from the $\gamma-$ray spectrum, with its 
associated $1\sigma$ error,  is propagated through as error bars that 
add in quadrature to the Poisson flux errors. 
  In addition to the formal systematic errors, one also has extrinsic
errors of uncertain magnitude stemming from the assumption 
   of one continuous power law over a broad spectral range. 
%
%
%
    For times
    close to $t_0$ that are of interest in 
this study, the exact value of $t_0$  determines  the logarithmic
decay slope.  In this work we take the same $t_0({\rm XRT}) = t_0({\rm BAT})
 = t_0({\rm trigger})$, the GRB trigger time.
%
 A summary of the BAT and XRT derived measurements is given in Table 1.

\begin{deluxetable}{lcccc}
\tablecaption{Summary}
\tablehead{
\colhead{Parameter} & \colhead{GRB/Value} & \colhead{Reference\tablenotemark{a}}
} 
\startdata
                                                     &   GRB050315                               &  \cr
\tableline
     $T_{50}$(BAT)                                  & $25\pm5$ s                                & 1\cr
     $T_{90}$(BAT)                                  & $96\pm10$ s                               & 1\cr
     fluence\tablenotemark{b}    (BAT)              & $3.4\pm0.3$                               & 1\cr
     $\Gamma$\tablenotemark{c}[1s peak] (BAT)  & $2.3\pm0.2$                               & 1\cr
     $\Gamma$[$T_{50}$] (BAT)                   & $2.02\pm0.07$                             & 2\cr
     $z$\tablenotemark{d}                         & $1.949$                                   & 3\cr
     $E_{\rm isotropic}$\tablenotemark{e}           & $3.2\times 10^{52}$                    &  \cr
     $\Gamma$[80s--300s]     (XRT)     & $2.5\pm0.4$                               & 2\cr
     $\Gamma$[300s--$10^4$s] (XRT)   & $1.7\pm0.1$                               & 2\cr
\tableline
                                                    &  GRB050319                                &  \cr
\tableline
     $T_{50}$(BAT)                                  & $124.1   \pm  0.4$ s                      &  \cr
     $T_{90}$(BAT)                                  & $141.2    \pm 0.8$ s                      &  \cr
     fluence   (BAT)                                & $1.6\pm0.2$                               & 4\cr
     $\Gamma$   (BAT)                               & $2.1\pm0.2$                               & 4\cr
     $z$                                            & $3.24$                                    & 5\cr
     $E_{\rm isotropic}$                            & $3.7\times 10^{52}$                    &  \cr
     $\Gamma$[90s--300s]     (XRT)   & $2.6\pm0.2$                               & 4\cr   
     $\Gamma$[300s--$10^4$s]  (XRT)   & $1.7\pm0.1$                               & 4\cr
\enddata
\tablenotetext{a}{References: 
(1) Krimm et al. 2005.
(2) V05.  
(3) Kelson \& Berger 2005. 
(4) C05.
(5) Fynbo et al. 2005.}
\tablenotetext{b}{$15-150$ keV, unit $=$ $10^{-6}$ erg cm$^{-2}$}
\tablenotetext{c}{Photon Index}
\tablenotetext{d}{Redshift}
\tablenotetext{e}{Isotropic Equivalent $\gamma-$ray energy, unit $=$ erg}
\end{deluxetable}


{\bf GRB 050315:}
 A detailed description of the XRT data reduction is given 
  in Vaughan et al. (2005 = V05).
The XRT count rate of GRB 050315 at the start of the pointed
observation was in excess of $100$~ct~s$^{-1}$ 
   ($\sim 3
\times 10^{-9}$~erg s$^{-1}$ cm$^{-2}$), resulting in heavy
pile-up in the PC-mode data. Ordinarily the XRT camera would
have switched to a different mode (e.g., WT or Photodiode
modes) in order to accommodate such a high rate, but
   the XRT was in Manual State at the time
of the trigger and remained in PC mode during the early
observations.

The most obvious effect of pile-up is an apparent loss of
counts from the center of the image, compared to the
expected Point Spread Function (PSF). This effect was used
to determine at what count rate pile-up can no longer be
ignored, by fitting the image radial profile with a 
    PSF model and successively ignoring the inner
regions until the model gave a good fit. The region over
which the PSF model gave a good fit is the region over which
pile-up may be ignored.  In the present analysis the central
$8$ pixels (radius) were ignored for (observed) count rates
between 1  and  5 ct s$^{-1}$, and the central $14$ pixels
(radius) were ignored for higher count rates. 
(Note one pixel corresponds to 2.36 arcsec.) 
  After
excluding the center of the image the fluxes were corrected
simply by calculating the fraction of the integrated PSF
used in the extraction.  (These results were obtained using
only mono-pixel events, i.e. {\tt grade = 0}, which should
be least affected by pile-up.)
A light curve was extracted over the $0.2-10$~keV band,
binned such that there were $25$ source events per bin, and
a background was subtracted using a large annulus concentric
with the source extraction region. Error bars were
calculated assuming counting statistics.

{\bf GRB 050319:}
 A detailed description of the XRT data reduction is given in C05.
The XRT count rate values were obtained extracting events ($0-12$ grade;
$0.2-10$ keV) in a circular region. Pileup in the first part of the 
observation was then corrected by excluding the central
pixels, fitting a PSF model
     to the wings of the emission, 
and rescaling the central portions using the instrumental
  PSF to recover the lost counts.
  Events were 
binned in order to have a constant S/N of 5. The light curve was then 
fitted with a broken power law with  two temporal
              breaks. The conversion factor 
from count rate to flux was obtained  by performing the spectral 
analysis of the whole XRT spectrum and by comparing the unabsorbed flux 
in the $0.2-10$ keV band with the average count rate in the same energy 
band. This correction factor was then applied both to
the XRT light curve and the best fit model.

\begin{figure}
\centering
\epsscale{1.15}
\plotone{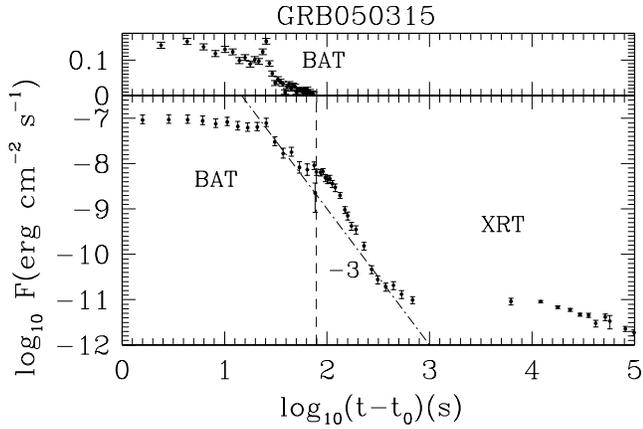}    
\caption{
The combined BAT/XRT $0.2-10$ keV light curve of GRB 050315.
   The small panel  on top
   shows the BAT data on a  log-linear scale, in units of background subtracted
15-350 keV flux per fully illuminated detector. The main, large panel
   shows the combined BAT and XRT data. The vertical dashed line shows the
   approximate time of the start of XRT observations, and the
   dot-dashed line indicates a logarithmic decay slope of -3.\label{Fig. 1}
}
\end{figure}

\begin{figure}
\centering
\epsscale{1.15}
\plotone{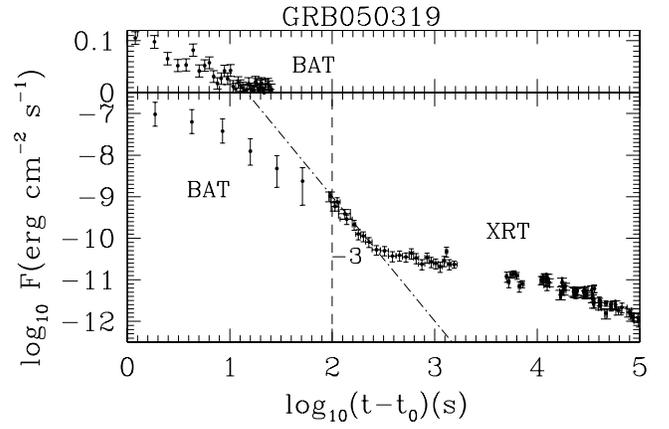}    
\caption{
The combined BAT/XRT $0.2-10$ keV light curve of GRB 050319.
    The conventions are the same as in Fig. 1.
                 \label{Fig. 2}
}
\end{figure}

 Figures 1 and 2  show the composite light curve decays for the $0.2-10$ keV fluxes,
extrapolated from the BAT and measured by   the XRT.  
    The dot-dashed line in each plot
    indicating a logarithmic slope of $-3$
    is not a fit to the data, but intended to be illustrative.
  Up to $\sim250$ s after burst onset,
                 one sees a 
  steep decay in the light curve.
  After this time the slope flattens abruptly,
   demarcating the time at which the prompt emission gives way to the
   early afterglow.
For  GRB 050315, 
exponential decays give a  better characterization 
   than a single  power law decay for the
BAT and XRT light curves for  $t-t_0 \la 300$ s.
The $e-$folding decay times are $\tau$(BAT) $\simeq 24 \pm  2$ s
 and  $\tau$(XRT) $\simeq 35 \pm  2$ s; after taking into account
the cosmological $(1+z)$ time dilation, these transform to
 $\tau$(BAT) $\simeq 8 \pm  1$ s and  $\tau$(XRT) $\simeq 12 \pm  1$ s 
   at $z=1.95$ (V05).
 This
slight difference between BAT and XRT is consistent with modest hard-soft
evolution.
 As discussed in detail in V05,
 the $t-t_0\la 10^3$ s XRT light curve for 
    GRB 050315 evolves through
flat $\rightarrow$ steep $\rightarrow$ flat phases 
   (followed by a second steepening seen
in later orbits). This first part of the light curve, until
the end of the steep descent at $\sim300$ s, can be modeled
either using a broken power law or an exponential decay.
%
%
%
%
 (The second break and
additional flat power law accounts for the
 true
afterglow emission.)
A single power law for the steep decay is not acceptable.
   The two
solutions are  (i)  a break in the power law from $n=2$ to $n=5$ at
$t-t_0\sim120$ s (V05, Table 2) or
(ii) 
an exponential decay. Both models give excellent
fits; formally the exponential model gives a worse $\chi^2$
  fit,  but has
two fewer free parameters.
It may be more appealing 
 due to its simplicity 
  than an arbitrary power law break.
 Exponential decays also avoid the problem of the choice of
$t_0$ which has a strong influence on the derived decay slope $n$.





\section {Discussion}


 We have presented  convincing evidence that for two GRBs observed by {\em Swift},
  the  prompt emission can be seen in X-rays up to about 300 s after the GRB
trigger.
     In addition, the  light curves from the BAT and XRT connect continuously,
   without there being a significant offset.
     For  completeness, we note that not all such GRBs for which
complete early-time  XRT observations exist share this property.
   For instance, Tagliaferri et al. (2005)  present data for two other GRBs,
    GRB 050126 and GRB 050219a,
      for
      which the early time XRT light curve lies significantly above an extension
   of the BAT $0.2-10$ keV (extrapolated) light curve.
  It is possible that strong spectral evolution,  and/or and non-power-law
spectral shape, may invalidate the simple prescription
 we and others have adopted of extrapolating the BAT flux into the XRT bandpass.
     Another possibility is that a flare occurred 
     in the X-ray bandpass (Burrows et al. 2005b),
     with the maximum located before the XRT observation began (i.e., at $t<t_0+100$ s).
     All five of the GRBs studied by Tagliaferri et al.
     show XRT light curves in which the initial steep decay gives way at later times
 to a more shallow decay, thereby supporting the idea of the initial X-ray flux
  as representing
 a continuation  of the prompt emission.
     Campana    et al. (2005) present an  XRT light curve for GRB 050128
   that shows evidence for flat decay at $t\la300$ s, followed by a steeper
decay out to $t\ga10^5$ s.
%
%
%
   It is difficult to form a general hypothesis of the early X-ray behavior
based on so few examples
(cf. Nousek et al. 2005),
    but it may be that for most GRBs
   the intrinsic tendency is for the prompt decay up to $\sim300$ s
  to be steep,
    as in GRB 050315 and GRB 050319, whereas for 
others a variety of systematic effects, such as viewing geometry, rapid cooling of the ejecta, 
    and evolutionary effects
      such as the shifting of  the
   synchrotron cooling frequency $\nu_c$
     out of the observational (XRT) bandpass,
   conspire to  distort 
    and hence obscure this
simple, underlying  behavior.

Within the theoretical framework of the expanding, relativistic blast wave model
  in which synchrotron emission from relativistic electrons dominates,
   the power law decay index for the decaying light curve depends only
           on the index of the power-law distribution
   of electrons
  with energy, the density stratification of the medium into which the burst propagates,
 and  the location of the frequency of the observing bandpass relative to
   $\nu_c$.
The most straightforward interpretation 
   of the steep initial decay
      for GRB 050315 and GRB 050319  may be the
 ``curvature effect''
associated with
 the  time delay 
 from high latitude emission within the relativistic ejecta.
This effect is due to the fact that, when the internal shocks stop radiating, 
   an observer viewing the emission close to the primary velocity vector
  of the ejecta  sees emission from
larger and larger viewing angles due to the Doppler delay effect
  (Kumar \& Panaitescu 2000=KP00, 
  Dermer 2004).

As noted in the previous section,  for GRB 050315 an exponential decay fits better
than a power law decay, indicating that at least one of the underlying assumptions
entering into the power law derivation  is not fulfilled.
%
%
%
%
%
    An exponential decay from the large-angle GRB emission would be obtained
    if the comoving frame energy band which is Doppler-shifted to the 
observer's $0.2-10$ keV
    band were above the cooling frequency only if the outflow were
tightly
   collimated, and we see its boundary. If
   the GRB 
emission
   stopped at $t_0$,
   then at  $t-t_0\sim100$ s, we see the emission 
from an angle
      $(100 \ {\rm s}/t_0)^{1/2} \gamma^{-1}$ ($< 2 \gamma^{-1}$)
                                           because the arrival time for the
   large-angle emission increases as the square of the angle
   from whence
that emission arises.
   Hence, the large-angle GRB emission would exhibit an 
exponential decay
   (above the cooling frequency) only if the jet is narrower than 1 degree.
  On the other hand, if the break in the XRT light curve at $t-t_0\simeq2\times 10^5$ s
represents the jet break, the observed $E_{\rm iso}$ value for GRB 050315
implies a jet opening angle $\theta_0\simeq 5^{\circ}$ (V05), which  would be
inconsistent with this explanation.
  One possible remedy may involve
   some aspect of
  alternative models
    that advocate  a much smaller beaming angle ($\la 1^{\circ}$)
   and larger Lorentz factor  ($\ga 10^3$) for the GRB jet
  (e.g., Dar \& R\'ujula 2004).

The transition at $t\simeq250-300$ s
   in our reference frame
       to a much flatter decay law in GRB 050315 and GRB 050319
   may provide a clue to the time scale for the relativistic shell to 
decelerate as it moves into the ISM gas.
KP00 give  the shell deceleration time, measured 
        in the local rest frame 
  at a given $z$, 
       as
   $100$ s $E_{52}^{1/3} (1-\eta)^{1/3} (\eta n_0 \gamma_2^8)^{-1/3}$,
   where $E_{52}$ is the isotropic equivalent $\gamma-$ray energy in units of 
$10^{52}$ erg, $\eta$ is the efficiency factor for converting internal energy
of the explosion into $\gamma-$ray energy, $\gamma_2 = \gamma_0/10^2$ is the 
initial Lorentz factor of the ejecta, scaled to 100,
       and $n_0$ is the number density of the ISM.
   (The deceleration time measured in the comoving ejecta frame
             is larger by a factor 
   $\sim2\gamma^2 \simeq 10^4$.)
     The times at which 
  the initial steep XRT decays abruptly give way to
   much shallower decays are
       $\sim100$ s in the frame of an observer at a cosmological redshift  $z=1.95$
for GRB 050315 
  ($\sim300$ s
 in our reference frame),
     and 
     $\sim60$ s at $z=3.2$
   for GRB 050319
    ($\sim250$ s in our frame).
   The fact that the time of our flattening is consistent with the 
theoretical 
     deceleration 
time  adds strength to the standard model of relativistic ejection and prompt emission,
followed by deceleration and afterglow emission.
%
%
   As a potential caveat to this interpretation,
  Zhang et al. (2005) carry out detailed numerical calculations
  of the curvature effect and find that
  the observed 
    transition time between steep and shallow decay
    may 
    only be an upper limit to the deceleration time. The
fireball
could  well be decelerated earlier, 
          but the deceleration signature
(marked by a rising phase followed by a $n\simeq-1$ decay) could be buried beneath
the steep-decay component. 
    Zhang et al. (2005)
 use  the observed transition times
   for GRB 050315 and GRB 050319
        to set lower limits on the
      initial
fireball 
Lorentz factors.


\section {Conclusion }

We present combined BAT/XRT data from two GRBs observed
by {\em Swift} for XRT observations began within 100 s of the 
BAT trigger. 
The data presented herein
give a clear indication that the prompt emission 
              and late afterglow emission are two
          distinct components. The early X-ray afterglow 
          is the tail of the prompt $\gamma-$ray emission,
                   and the late X-ray
afterglow is the normal forward shock afterglow. 
  This lends support to the prevailing notion that prompt emission
is from internal shocks rather than external shocks.

\end{document}